# Investigating students' gender expression and its relation to sense of belonging in introductory physics courses

Noah Leibnitz and Yangqiuting Li
*Department of Physics, Oregon State University, Corvallis, Oregon 97331, USA*

## Abstract

Despite ongoing efforts to promote diversity and inclusion, women and gender minorities remain underrepresented in physics. Research aiming to understand how physics learning environments contribute to these trends has increasingly attended to students' experiences of recognition and belonging. Gender expression offers a valuable lens for this work because it captures the nuanced and context-dependent ways in which gender is socially enacted and interpreted within cultural contexts. In our recent quantitative work, we adapted a gradational metric, previously developed in sociological research, to examine students' gender expression in introductory undergraduate physics courses. We found that students often reported discrepancies between self- and reflected appraisals (i.e., how they view themselves and how they believe they are viewed by others) of masculinity, femininity, and androgyny, and that these discrepancies were correlated with their sense of belonging in physics. In the present study, we qualitatively examine the sources of students' self-reflected gender appraisal discrepancies (i.e., differences between self- and reflected appraisals) and how these discrepancies relate to their sense of belonging in introductory physics courses. Through semi-structured interviews with 26 students from various gender identity groups, we identified two common ways students explained their gender appraisal discrepancies: they felt that others, including—and for some interviewees, especially—their physics peers, might not know them well on a personal level and might hold different views on gender than they did. Students' accounts further suggested two ways these discrepancies were connected to belonging: both discrepancies and lower belonging were linked to feeling not personally known by their physics peers, and many interviewees also described pressure to alter their gender expression in physics to fit in, gain recognition, or establish their presence. Together, these findings suggest that gender appraisal discrepancies may reflect broader experiences of misrecognition, disconnection, and self-presentational pressure in physics learning environments. Supporting students' diverse and authentic self-expression and fostering interpersonal connections among peers may therefore be important steps toward cultivating a stronger sense of belonging in physics.

## I. Introduction

Physics continues to face persistent gender disparities in participation and persistence. For example, although women earn approximately 60% of bachelor's degrees in the U.S., only 25% of the physics undergraduate degrees are earned by women [1,2]. Importantly, research shows that even when women perform as well as or better than men, they are more likely to leave physics, suggesting that factors beyond academic performance influence persistence in the field [3].

Beyond participation, prior research has demonstrated inequities across gender identity groups in academic outcomes and motivational beliefs [4]. Studies have also documented experiences of marginalization and exclusion reported by women and gender minority students in physics learning environments, which often contribute to decisions to leave the field [5–9]. Collectively, these findings highlight ongoing challenges related to gender equity and inclusion in physics and have drawn substantial research attention to understanding and promoting more inclusive learning environments.

Much of the existing research on gender equity in physics has commonly examined gender as a categorical identity variable. While this work has been critical for documenting inequitable outcomes across groups, categorical approaches often focus on group-level patterns and are therefore less well suited for examining how gender is enacted and negotiated within specific classroom contexts [10]. Such dynamics may be better understood by attending to students' gender expression [11]. Gender expression refers to how individuals present and enact gendered selves in interaction, reflecting how gender is socially performed, perceived, and interpreted in various contexts [11,12]. In physics education research, attending to students' gender expression has helped illuminate forms of marginalization that women and gender minority students may experience [13–15]. For example, women in physics may feel pressure to distance themselves from stereotypically feminine traits to fit in with predominantly "masculine" modes of expression [13–15]. Additional research shows that gender minority students also experience pressure to conform or conceal aspects of their gender expression to avoid exclusionary behaviors such as misgendering, harassment, or social isolation [9,16,17]. Collectively, these studies suggest that students' experiences of gender expression are important for understanding processes of inclusion in physics learning environments [5–10,18].

Building on this literature, our recent quantitative work sought to examine students' self-appraisals and reflected appraisals of femininity, masculinity, and androgyny as one way to study gender expression in introductory physics courses. We adapted a gradational metric of gender expression appraisals, previously employed in other contexts, and used it alongside gender identity to examine associations between students' gender expression appraisals and social psychological factors such as sense of belonging [28,29]. Within gender identity groups, we found substantial variation in students' self-appraisals of femininity, masculinity, and androgyny. Additionally, students' self-appraisals (how they viewed themselves) did not always align with their reflected appraisals (how they believed others viewed them). In examining these self-reflected appraisal discrepancies, we found that particular discrepancy patterns were correlated with students' sense of belonging in the physics class. These findings suggested that mismatches between students' self-appraisals and reflected appraisals may capture aspects of their social experiences in physics that are relevant to belonging. However, quantitative work alone cannot fully account for the meanings and experiences underlying these patterns. Further qualitative inquiry is needed to better understand how students interpret gender expression survey items, how they explain differences between their self- and reflected appraisals, and how these discrepancies relate to students' sense of belonging in physics.

In the present study, we build on this prior work by employing a qualitative methodology to investigate students' self- and reflected appraisals of gender expression in physics and their relationship to sense of belonging. Our primary objectives were twofold: to inform the interpretation and future application of gender expression metrics in physics contexts, and to explore potential explanations for the observed self-reflected appraisal discrepancies and their relationships with belonging. To this end, we interviewed students of various gender identities enrolled in introductory physics courses at a large public university in the Northwest United States. Interviews and subsequent analyses focused on the students' understandings of feminine, masculine, and androgynous expressions, their reasoning about their responses to the gradational gender expression metric, and their perspectives on discrepancies between self- and reflected appraisals and on the experiences that shape their sense of belonging.

In the next section, we provide background for this study, including a review of relevant literature on gender expression, sense of belonging, and their intersections within physics education. Before proceeding, it is important to clarify the terminology used in this study and the reviewed studies. The following definitions will be used:

- ***Gender*** is a social construct that encompasses the roles, behaviors, expectations, and identities associated with femininity or masculinity within a particular society or culture. Gender exists along a spectrum rather than being binary and can vary across cultures and evolve over time.
- ***Sex*** is a set of biological and physiological traits (e.g. anatomy, hormones) that is typically assigned at birth as male, female, or intersex [30].
- ***Femininity*** is a set of attributes, behaviors, and roles generally associated with women and girls. Femininity can be understood as socially constructed.
- ***Masculinity*** is a set of attributes, behaviors, and roles generally associated with men and boys. Masculinity can be understood as socially constructed.
- ***Androgyny*** has often been described as involving both feminine and masculine traits or characteristics, although this construct has been defined in varying ways and remains contested [31,32].
- ***Fluidity*** refers to changes in a person's gender expression over time or depending on the situation [33].
- ***Self-appraisal of femininity, masculinity, or androgyny*** refers to an individual's self-perception of their own degree of femininity, masculinity, or androgyny.
- ***Reflected appraisal of femininity, masculinity, or androgyny*** refers to how individuals perceive their femininity, masculinity, or androgyny to be viewed by others [34,35].
- ***Self-reflected gender appraisal discrepancy*** refers to differences between an individual's self- and reflected appraisals of femininity, masculinity, or androgyny. As described further in the methodology section, we constructed a numerical measure for appraisal discrepancies by subtracting respondents' reflected appraisal scores on the gradational gender metric from their self-appraisal scores.

## II. Background

### A. Gender inequities in physics learning environments

Gender inequities persist across multiple dimensions of physics education. Prior research has documented such inequities in academic outcomes [4,36–39], motivational beliefs such as self-efficacy and interest [22,26,27,40–45], and students' experiences of physics learning environments [4,22,23,26,46,47]. For example, compared to men, women often report lower perceived inclusiveness of physics learning environments, including perceived recognition from others, peer interaction, and sense of belonging [46,47]. Together, these findings suggest that gender inequities in physics extend beyond achievement gaps and are closely tied to students' experiences within learning environments.

A growing body of research points to the role of disciplinary culture in shaping students' experiences and perceptions of inclusion [4,48,49]. Prior studies have shown that women often experience physics learning environments and the discipline more broadly as unwelcoming and exclusionary [50–54]. Such perceptions may stem from overt forms of discrimination, such as harassment [55], but also from subtle yet persistent forms of exclusion, including microaggressions—everyday slights or indignities, whether intentional or unintentional, directed at individuals based on their gender [5,17]. Notably, experiences of feeling unwelcome and excluded have been linked to women's decisions to disengage from or leave the field [23,56,57].

Research on gender minority students' experiences in physics, together with findings from broader STEM contexts, suggests that these students face similar challenges in learning environments. For example, gender minority students (e.g. transgender and gender nonbinary students) often report lower levels of belonging and less positive perceptions of STEM learning environments [17,58]. In physics specifically, Barthelemy et al. found that gender nonconforming respondents reported higher rates of discomfort in their physics classes, departments, and workplaces, and were significantly more likely to report experiencing or observing exclusionary behavior than other respondents [8]. These exclusionary behaviors included being shunned, ignored, harassed, and misgendered. Further studies suggest that these exclusionary climates influence students' decisions about persisting in physics and related fields [6,59].

Many of these exclusionary dynamics are closely tied to gendered stereotypes in physics. A particularly relevant stereotype involves the association of physics ability with brilliance, which itself is often stereotypically associated with men [60–62]. The belief that success requires a fixed, "you-either-have-it-or-you-don't" type of intelligence has been shown to impact women's and other marginalized students' sense of belonging, interest, and mindset in physics and related STEM fields [19,24,25,41,63–66]. Additionally, several studies have noted that underrepresentation of women is most prevalent in disciplines, such as physics, where innate ability is stereotypically believed to be essential for success [67,68]. This research suggests that gender interacts with disciplinary culture and cultural stereotypes to influence students' experiences in physics learning environments.

### B. Gender expression in physics learning environments

Approaching gender through the lens of gender expression involves attending to the fluid and context-dependent ways in which gender is enacted and negotiated within specific learning contexts. Research using this perspective highlights how gender permeates students' experiences and interactions in educational settings. Across educational contexts, studies have consistently shown that when an individual's gender expression is validated and supported, it fosters a sense of integration and safety, whereas its restriction can lead to marginalization and disengagement [69,70].

In physics, research on gender expression suggests that dominant gendered norms and stereotypes within the discipline may contribute to tensions related to gender expression for many women and other gender-minoritized individuals. These tensions may shape how individuals experience and navigate the discipline. For example, several studies have found that women in physics frequently engage in complex negotiations of gender expression stemming from perceived incompatibilities between their "woman" and "physicist" identities [13–15,54,71,72]. Gonsalves et al. argued that the masculine connotations associated with physics (often construed as gender-neutral) constrain the range of recognizable identity positions within the discipline [15]. Drawing on case studies of undergraduate and graduate students as well as research scientists, they and others have documented how women physicists, in negotiating these limited identities, frequently distance themselves from traditionally conceptualized femininity, performing instead what Halberstam terms "female masculinity" [13–15,72–74].

Further research has examined how such "distancing" is enacted in practice. These studies describe several forms of identity negotiation through which women seek recognition and belonging in masculine-dominated environments. Tsai, for example, found that women in physics were often positioned—and sometimes positioned themselves—as "exceptional" rather than representative of women more broadly, reflecting the contested discursive terrain of gender and physics. Related forms of identity negotiation may fall into the categories that Lugones refers to as "fragmentation" and "multiplicity" [75]. Drawing on Lugones's work, Ong describes fragmentation as the "process of temporarily splitting oneself to minimize cultural differences between oneself and other members of a community" and multiplicity as the simultaneous and unapologetic "occupation of multiple, but sometimes competing, identities and memberships" [54]. Ong observed that women of color employed both such strategies as they navigated recognition and belonging by approximating the "ordinariness" of white men in physics [54].

Similarly, studies have shown that gender minorities face persistent challenges related to gender expression in physics environments. For example, Barthelemy [9] found that expressions of gender were among the most frequent targets of exclusionary behavior reported by LGBT physicists, with reported forms including misgendering, derogatory comments, and social exclusion. Barthelemy [9,76] also found that both transgender and cisgender physicists who presented in gender-nonconforming ways experienced harassment and exclusion tied to their appearance and expression. Collectively, these studies suggest that examining how gender is socially constructed and enacted within physics contexts can provide important insight into how disciplinary cultures impact individuals' participation.

### C. Prior quantitative studies of gender expression in physics

Building on the growing recognition that gender expression plays an important role in shaping individuals' social experiences, recent research has begun to operationalize gender expression as an analytical construct, offering quantitative approaches for examining these dynamics. Magliozzi et al., for example, introduced a gradational approach to measuring gender expression appraisals, asking respondents to rate their level of masculinity and femininity on a Likert-type scale [77]. Importantly, this metric includes two components: one that asks respondents to rate how they view themselves (*self-appraisal*) and another that asks how they believe others perceive them (*reflected appraisal*) [77] (fig. 1). In a later study, androgyny was added as another dimension [26], further broadening the scope of this measure. While the metric developed by Magliozzi et al. has been adapted to investigate a range of topics [77–81], its use in physics education research remains limited.

In our recent studies, we adapted this gradational gender metric to the context of physics education to examine students' self- and reflected appraisals of femininity, masculinity, and androgyny within physics learning environments. In our first study using this metric, we observed substantial variation in students' self-appraisals of gender expression, with ratings of masculinity, femininity, and androgyny spanning the full range of possible responses across gender identity groups [28]. Additionally, a significant proportion of students reported discrepancies between their self- and reflected appraisals of gender expression across all three dimensions of the metric [28]. Notably, some of these self-reflected appraisal discrepancies correlated with students' sense of belonging; in particular, negative femininity discrepancy, meaning that students believed others perceived them as more feminine than they perceived themselves, was associated with lower sense of belonging [28].

To examine whether these patterns would recur across institutions, we subsequently conducted a cross-institutional replication at the institution where the present qualitative study was conducted. Results from this replication study aligned with the previous findings: there was substantial variation in students' self-appraisals of gender expression, and self-reflected appraisal discrepancies correlated with sense of belonging [29]. Specifically, we observed that negative self-reflected appraisal discrepancies in femininity continued to be associated with lower sense of belonging. Additionally, we found that positive masculinity discrepancy, meaning that students felt their masculinity was underrecognized by others, was also associated with lower levels of belonging (Fig. 2) [29]. The consistency of results across institutional contexts suggested that these patterns may reflect broader features of physics learning environments rather than purely local conditions.

These findings suggest that misalignments between students' self- and reflected appraisals of gender expression are associated with their sense of belonging in physics. However, quantitative analyses alone cannot fully explain how students interpret these appraisals, how such discrepancies arise, or how they are experienced in practice. Because belonging plays a central role in students' academic experiences and persistence [22,23,26,82,83], these patterns underscore the need to better understand how gender expression and belonging intersect within physics contexts.

(a) In general, how do you see yourself? (Please answer all three scales).

| | Not at all | 1 | 2 | 3 | 4 | 5 | Very |
| --- | --- | --- | --- | --- | --- | --- | --- |
| Feminine | O | O | O | O | O | O | O |
| Masculine | O | O | O | O | O | O | O |
| Androgynous | O | O | O | O | O | O | O |

(b) In general, how do most people see you? (Please answer all three scales).

| | Not at all | 1 | 2 | 3 | 4 | 5 | Very |
| --- | --- | --- | --- | --- | --- | --- | --- |
| Feminine | O | O | O | O | O | O | O |
| Masculine | O | O | O | O | O | O | O |
| Androgynous | O | O | O | O | O | O | O |

**Fig. 1.** The gradational gender metric used in our prior quantitative studies [28,29]. The metric was also used in both the survey and interview stages of the present study. Participants' responses to the set of items for question (a) constitute their self-appraisals of femininity, masculinity, and androgyny, respectively, while those for question (b) constitute their reflected appraisals. Differences between corresponding responses on the two question sets indicate a self-reflected gender appraisal discrepancy.

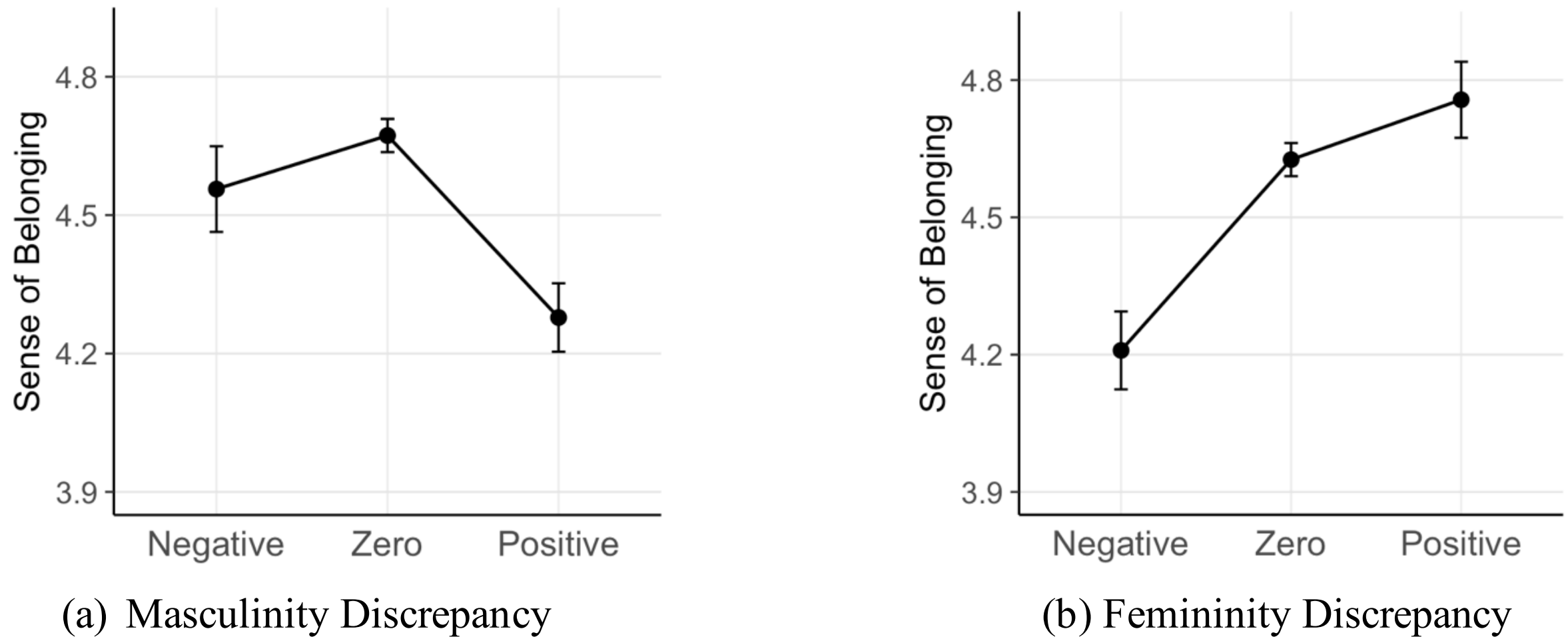

**Fig. 2** Key results from the replication study conducted at the institution where the present qualitative study took place. Students with positive masculinity discrepancy (i.e., those who believed that others viewed them as less masculine than they viewed themselves) reported lower average levels of belonging in their physics class than those with either negative or zero masculinity discrepancies. A comparable pattern was observed for femininity, but in the opposite direction. Students with negative femininity discrepancy (i.e., those who believed that others viewed them as more feminine than they viewed themselves) reported lower average levels of belonging than those with either positive or zero femininity discrepancies.

### D. The role of belonging in learning, equity, and inclusion

In education research, disciplinary sense of belonging has been commonly defined as "the extent to which students subjectively perceive that they are valued, accepted, and legitimate members in their academic domain" [42]. This construct builds on foundational work in psychology, where belonging has long been recognized as a fundamental human need [86,87].

In physics, sense of belonging has received significant research attention and been linked with various aspects of students' physics learning. For example, previous studies show that students' sense of belonging in introductory physics courses predicts both their course grades and their scores on the Force-Concept Inventory [22,26]. Prior studies have also found that sense of belonging predicts students' self-efficacy and persistence intentions in physics [23,82,83]. Moreover, additional research has identified sense of belonging as a significant predictor of physics identity, a construct that plays a central role in shaping students' aspirations to pursue physics-related careers [27].

Prior studies have revealed gender inequities in students' sense of belonging. For example, women often report a lower sense of belonging in physics classes compared to men [22–24,26,88]. Some research suggests that the gender composition of the learning environment can influence students' sense of belonging [89]. However, other research shows that this disparity persists even in courses where women are well represented or constitute the majority, indicating that numerical representation alone cannot fully explain gender differences in belonging [4,26]. Additionally, prior research suggests that gender disparities in belonging are consequential because sense of belonging has been linked to persistence in physics and other STEM fields, and some studies indicate that belonging may be a stronger predictor of persistence for women than for men [4,21,24]. Although fewer studies have specifically examined gender minority students' sense of belonging in physics, research in broader STEM contexts has documented that they often report lower levels of belonging and less positive perceptions of STEM environments [17,58].

## III. The present study

The present qualitative study builds on our prior quantitative work, which showed that students' self- and reflected appraisals of masculinity, femininity, and androgyny did not always align and that such discrepancies were often associated with students' sense of belonging [28,29]. Specifically, this study took place at the same institution as our previous replication study, in which both positive self-reflected appraisal discrepancies in masculinity and negative discrepancies in femininity correlated with lower belonging. To better understand how appraisal discrepancies emerge in physics learning environments and how they relate to students' sense of belonging, we conducted in-depth, semi-structured interviews with students who, at the time of the study, were enrolled in an introductory physics course. The following research questions guided this study:

**RQ1:** What are physics students' understandings of masculinity, femininity, and androgyny, and how do they place their self- and reflected appraisals on the gradational gender metric?

**RQ2**: What do physics students cite as the sources of their appraisal discrepancies (i.e., differences between self- and reflected appraisals)?

**RQ3**: How do students' experiences of gender expression and appraisal discrepancies in physics learning environments relate to their sense of belonging?

## IV. Theoretical framework

We adopted Judith Butler's theory of gender performativity, which emphasizes the socially constructed nature of gender, as the theoretical framework for this study [11]. From this perspective, gender is not an innate trait or stable essence, but is constituted through repeated, socially mediated acts that become recognizable as "masculine," "feminine," or otherwise gendered within particular cultural contexts. As Butler expounds in an essay concerning the theory, "there is neither an 'essence' that gender expresses or externalizes nor an objective ideal to which gender aspires," and therefore "the various acts of gender create the idea of gender… without [them] there would be no gender at all" [90]. Gender, in short, involves the ongoing interpretation of repeated performances (e.g., actions, gestures, speech acts) that both draw on and reinforce existing gender norms. Because such interpretations may vary over time and across social environments, gender is both temporally and culturally situated, constantly reconstructed and renegotiated. Therefore, the theory of gender performativity provides a productive framework for examining how students navigate and negotiate gender expressions within physics learning environments. In physics classrooms, where gender is performed and interpreted within particular institutional and cultural contexts, these processes of recognition may shape students' experiences of learning and belonging.

Importantly for this study, these perspectives underlie the premise that gender can be interpreted differently by individuals, who may themselves be aware of these differences. An individual's understanding of their gender performance may not align with how others interpret it; the "doing" of gender, as Westbrook and Schilt note, lends itself to various possible "determinations" of gender [91]. Moreover, prior research has shown that discrepancies between how people perceive their own gender and how others interpret it are associated with negative health outcomes, including heightened stress [79,92]. Such findings motivate our focus on gender appraisal discrepancies. In our prior quantitative work, we operationalized these discrepancies using a gradational gender metric that distinguishes between students' self-appraisals (how they view themselves) and reflected appraisals (how they believe others view them). In the present qualitative study, we build on that work by examining how differences between students' self- and reflected appraisals arise and are experienced within physics learning environments.

## V. Methodology

### A. Positionality Statement

We approach this work as physics education researchers whose scholarly and personal experiences inform our interest in issues of gender, inclusion, and belonging in physics learning environments. Both authors are trained physicists with experience teaching and conducting research in undergraduate physics contexts at a large public research university in the United States. The first author identifies as a White, cisgender man and is a graduate student working toward a Ph.D. in physics education. The second author identifies as an Asian cisgender woman and has extensive research experience in gender equity, sense of belonging, and inclusiveness in physics learning environments. These experiences shaped our attention to gendered dynamics in physics classrooms and motivated our use of gender expression as a research lens.

### B. Participants

The participants of this study were 26 undergraduate students enrolled in introductory physics courses at a large public research university during the Spring or Fall 2024 terms. The sample included students from both algebra- and calculus-based introductory physics courses—representing a range of majors in engineering and the sciences—as well as students from various gender identities and racial/ethnic backgrounds. A detailed breakdown of participant demographics by specific identity categories is shown in Table 2 in the Results section.

Participants were recruited through a combination of convenience and criterion sampling [84]. Instructors of all introductory physics courses offered during the Spring and Fall 2024 terms were asked to share study invitations with their students, and interviewees were selected from those who expressed interest. When making such selections, we intentionally included students from diverse demographic groups, seeking to include representation from each gender and racial/ethnic group represented among interested respondents. In addition, all students in the introductory physics courses completed a pre-course survey that included the same gradational gender expression metric used in our quantitative studies [28,29]. In alignment with the study's focus, we prioritized interviewing students who reported discrepancies between their self- and reflected appraisals of gender expression.

Participants' race/ethnicity was collected using an open-ended survey item ("What race and/or ethnicities do you identify with?"). Gender identity was collected using the survey item, "What is your current gender identity? (check all that apply)," with response options: man, woman, transgender, cisgender, nonbinary, gender fluid, agender/I do not identify with any gender, prefer not to state, and gender not listed (my gender is ___). Participants could select multiple options for gender identity and could also provide additional text in the free-response field, which allowed participants to describe their identities in their own terms rather than being limited to the listed response categories. Because participants could select any combination of gender identity terms, some participants identified themselves only as woman or man without additionally selecting terms such as cisgender or transgender. We therefore report participants' identities as self-described and do not infer identities that were not explicitly indicated.

### C. Course contexts

Participants were recruited from both the algebra-based and calculus-based introductory physics sequences. The algebra-based physics course sequence typically enrolls students majoring in the life sciences and related disciplines. The calculus-based sequence typically enrolls students pursuing degrees in engineering, physics, and other physical sciences. Both sequences are offered within a quarter-term system. Courses in the algebra-based track are each 5 credits, whereas the calculus-based courses are 4 credits. All courses meet for approximately 3 to 4 hours of lecture per week and include a weekly laboratory session lasting 2 hours. Laboratory sections are capped at 30 students, who work in groups of 3-4 students, while lecture sections typically enroll between 100 and 200 students. Both sequences incorporate a variety of active learning components and emphasize multiple representations, modeling, and sensemaking. The first course in each sequence covers Newtonian mechanics, where students build and apply models of motion, forces, energy, and momentum. The second course extends these models to topics such as rotation, oscillations, thermodynamics, and optics. The third course introduces electricity and magnetism, focusing on electric and magnetic fields, circuits, and induction.

### D. Data Collection: Semi-Structured Interviews

Interviews followed a protocol that was developed around the following areas of focus: 1) understandings of femininity, masculinity, and androgyny, 2) rationale for ratings on the gradational gender metric, 3) perceived sources of self-reflected appraisal discrepancies, and 4) experiences of gender expression and sense of belonging in physics learning environments. Throughout each interview, the interviewee's responses to the gradational gender metric were used to elicit deeper discussion about their gender expression in physics environments and their experiences related to self-reflected appraisal discrepancies. We note that, while the survey items that constituted students' reflected appraisals asked how respondents believed they were viewed by "others" in general, in their interviews students were also asked specifically about how they believed they were viewed by their physics peers. Example questions from the interview protocol are shown in Table 1. The protocol was developed in alignment with the research questions through an iterative process of refinement. An initial draft was used to conduct pilot interviews with three graduate students and two faculty members to evaluate the clarity, relevance, and sequencing of the questions. Feedback from these pilot interviews was used to refine wording, eliminate redundancies, and ensure that each question effectively elicited responses aligned with the study's research aims. This process also helped identify areas where follow-up prompts or transitions were needed to improve conversational flow and participant engagement. The protocol was then used during 13 interviews that were conducted in the Spring term of 2024. Small refinements were made after this first round of interviews to further improve clarity and flow, and an additional 13 interviews were conducted during the Fall term of 2024.

**Table 1.** Example questions from the interview protocol

**Understandings of masculinity, femininity, and androgyny**

1. When you hear the terms masculinity, femininity, and androgyny, what do they mean to you?

**Rationale for ratings on the gradational gender metric**

2. Can you share why you rated your masculinity/femininity/androgyny as ___? What traits or characteristics about yourself did you consider when making those ratings?
3. What leads you to think that others would rate your masculinity/femininity/androgyny as ___? What traits or characteristics do you think others consider when making such evaluations?

**Perceived sources of appraisal discrepancies**

4. Why do you think others might see your masculinity/femininity/androgyny differently from how you see yourself?
5. Would you prefer that others perceive your masculinity/femininity/androgyny to be more aligned with how you view yourself? Why or why not?
6. How would you compare how people in your physics class versus people outside of your physics class perceive your masculinity, femininity, and androgyny? What makes you think so?

**Experiences of gender expression and sense of belonging in physics learning environments**

7. How do you usually express your gender in physics learning environments (such as lectures, labs, recitations, office hours, learning groups)?
8. Do you feel a sense of belonging in your physics course? Can you share moments when you felt included and part of the class, as well as times when you did not?
9. Have there been moments in your physics courses when you felt included or excluded because of your gender expression? If so, can you describe those experiences?

Prior to each interview, participants were sent a document outlining information necessary to provide informed consent, including the purpose of the study, the voluntary nature of participation, and potential risks and benefits of participation [93]. To ensure properly informed consent, participants were reminded before each interview of the voluntary nature of their participation, informed of the measures that would be taken to protect their anonymity, and provided an opportunity to ask clarifying questions. Prior to recording, the participants were also asked to complete a paper version of the gradational gender metric and to provide any information they felt comfortable disclosing about their gender and racial/ethnic identities. This step was taken both to accurately record participants' identities and to provide a reference for both the interviewer and the interviewee when discussing the rationale for the ratings provided. The interviews each lasted approximately 1 hour, at the end of which the interviewee was thanked and compensated for their time.

### E. Data Analysis

Audio recordings of the interviews were transcribed, and the transcripts were subsequently uploaded to the qualitative analysis software NVivo. Coding followed an iterative and collaborative process involving multiple coding cycles. In the first round of coding, the two authors independently coded a subset of the data using a combination of inductive codes and deductive codes informed by the research questions and theoretical framework. The authors then met to resolve discrepancies between their two sets of codes and develop an integrated codebook. The first author used the integrated codebook to code the remaining transcripts. Inductively generated codes were occasionally added through consultation with the second author. In the second round of coding, initial codes were refined and consolidated into larger analytic themes. "Pattern codes" were also used to categorize quotes that evidenced relationships between the emerging themes [94]. Second-round coding was conducted primarily by the first author. However, both authors met regularly to discuss the coding process, and substantive revisions to the codebook were made collaboratively.

Additionally, numerical self–reflected appraisal discrepancies in femininity, masculinity, and androgyny were calculated for each participant. Appraisal discrepancies were defined as the difference between each participant's self- and reflected appraisals on the gradational gender metric, calculated as self-appraisal minus reflected appraisal. For example, an appraisal discrepancy in femininity of +1 indicates that the participant viewed themselves as more feminine than they believed others perceived them. Likewise, a discrepancy in masculinity of -1 indicates that the participant believed others viewed them as more masculine than they saw themselves. The magnitude of discrepancy reflects the degree of misalignment between self- and reflected appraisals, with larger absolute values indicating greater perceived difference.

## VI. Results

In this section, we present our interview findings in three sections corresponding to each research question: (i) Student understandings and uses of the gender expression terms "masculinity, femininity, and androgyny"; (ii) Sources of self–reflected gender appraisal discrepancies; and (iii) Connections between self–reflected gender appraisal discrepancies and sense of belonging in physics. Each section includes illustrative quotes from participants across demographic groups and highlights how these perspectives help illuminate the complex ways gender expression intersects with students' experiences in physics learning environments.

### A. Student understandings and uses of the gender expression terms "masculinity, femininity, and androgyny" (RQ1)

#### 1. The role of social stereotypes in students' conceptions of gender

For every interviewee, the terms masculinity, femininity, and androgyny conjured images and characteristics that aligned with common social stereotypes. When asked to define or describe each term, the interviewed students often referred to examples of masculinity and femininity that were related to one's appearance, personality, and hobbies. For example, Wood said that

> Wood: *…my mom was more feminine, and she was a little more delicate… and the way she approached things was a lot gentler. For example, she would appeal more to emotion. But my dad would appeal more to logic…*

Fisher shared a related conceptualization of masculinity and femininity, focusing more on externally observable characteristics:

> Fisher*: Feminine presenting, on the outside, would be kind of more standard stuff where it's like long hair, and then the more standard like what a girl would look like…and then more masculine stuff is associated with shorter hair, facial hair, or stuff like that.*

All interviewees drew on common social stereotypes when asked to share their understanding of the terms masculinity and femininity, and many spoke similarly to Wood and Fisher. Roughly two-thirds of the interviewees associated femininity with being gentle, appealing to emotion, and appearing delicate or pretty. Conversely, they often associated masculinity with use of logic, being less emotionally expressive, and appearing strong. The traits and characteristics students most frequently associated with each category aligned with gendered traits operationalized in the Bem Sex-Role Inventory [95].

Similarly, in explaining their self-ratings on the gradational gender metric, students often drew on socially constructed notions of gender expression, referring to their appearance, personality, and behavior to describe how they were aligned with "masculine" or "feminine" traits. Sarah, for example, when asked to explain her ratings on the scale, recalled her descriptions from earlier in the interview, saying

> Sarah: *... I definitely present in a very feminine way. I put an effort into what I'm wearing ... always checking in on all my friends, making sure to ask how they're doing, and kind of just expressing more of those feminine traits that I kind of tried to define earlier.*

Some participants acknowledged their use of generalizations, noting that they drew on "typical" or "general" characteristics associated with masculinity and femininity. Others explicitly voiced their use of stereotypical social constructions. These comments were exemplified by Emi, who said

> Emi*: I feel like I like more things that fit kind of the general stereotype of femininity, like I have a pink water bottle…I just wear like clothes and stuff that look feminine…I guess fit the stereotype of that.*

Students generally interpreted androgyny through the lens of masculinity and femininity. Most students (17/26) defined androgyny as a blending of characteristics that were both feminine and masculine, while some students (11/26) conceived of being androgynous as having neither masculine nor feminine traits. We note that there was overlap between these two groups such that (7/26) students spoke simultaneously about androgyny being both a combination and a lack of masculinity and femininity. This suggests that these students' conceptualizations of androgyny were complex. A subset of students (5/26) were unfamiliar with the term androgyny.

### 2. Complexity and nuance in students' conceptions of gender

Although many students drew on common social stereotypes to describe their gender expression, a substantial number (17/26) also articulated more nuanced and complex understandings of gender expression, which also shaped their appraisals on the gradational gender metric. For example, some interviewees (8/26) expressed a belief that stereotypes did not completely encapsulate their conceptions of masculinity, femininity, and androgyny. Precisely how those understandings transcended stereotypes, however, was often too nuanced to articulate in general terms. Monica, for example, commented on the limitations of stereotypical definitions, saying

> Monica: *... there's ways to be masculine without being tough or rugged... And there's ways to be feminine without being cute and delicate. So, it's really difficult to define those two terms as anything other than, like, vibes I get on an individual person...*

Monica's comments suggest that their understanding of masculinity and femininity incorporates—but is not fully constituted by—stereotypes. This perception influenced how Monica made gender appraisals. Rather than systematically comparing traits against stereotypical conceptions, they interpreted others' gender based on a "vibe" that they described as difficult to define. Other students echoed Monica's sentiments. They often observed that their sense of gender derived from a vague or highly internalized "vibe," "feeling," or "energy."

Students' understanding of the complexity of gender was also reflected in their views on the relationship between masculinity and femininity. Although some students acknowledged that they often thought of masculine and feminine traits as opposites, a majority (20/26) of interviewees felt that masculinity and femininity were not mutually exclusive. Jaycee, for example, noted that

> Jaycee: *...I feel like you can do both and feel both. It doesn't really matter what characteristics there are...I definitely feel like you can be fully masculine and fully feminine.*

Marie added a more context-sensitive view:

Marie: *...there definitely can be that dichotomy there... especially when we're talking about the external characteristics... But when you're looking at both internal and external characteristics, I feel like someone could possess both masculine and feminine traits.*

Relatedly, most interviewees (15/26) expressed an understanding of gender fluidity. Phoebe typified this set of responses, saying

Phoebe: *I think people can identify as both [masculine and feminine] at the same time, like gender fluid folks who may one day want to wear a pretty dress and then the next day, they're like, 'I wanna wear a button-down.' So maybe it depends on the day, depends on vibe, too...*

In summary, students' understandings and uses of the gender expression terms were shaped by both societal stereotypes and personal, internalized conceptions. All interviewees drew on social stereotypes when defining the terms or explaining their ratings on the gradational gender metric. At the same time, many students acknowledged the limitations of these stereotypes and expressed more nuanced and complex interpretations of gender expression. Most students did not view masculinity and femininity as strictly opposed; instead, they emphasized fluidity and coexistence between masculinity and femininity. These findings not only illuminate the ways students navigate gender expression but also provide important context for: 1) interpreting how students perceive alignment or misalignment between their self- and reflected appraisals (RQ2), and 2) elucidating how students' experiences of gender expression and self-reflected appraisal discrepancies in physics learning environments relate to their sense of belonging (RQ3).

**Table 2.** Summary of participants' self- and reflected appraisals for masculinity (M), femininity (F), and androgyny (A) using the gradational gender metric, along with self-reported gender identities and races/ethnicities. Appraisal discrepancies represent the difference between each student's self- and reflected appraisals; positive values indicate higher self-appraisals than reflected appraisals, while negative values indicate the opposite.

| Pseudonym | Self-Appraisals (M, F, A) | Reflected Appraisals (M, F, A) | Appraisal Discrepancies (M, F, A) | Gender Identity/ies | Race(s)/Ethnicities |
|---|---|---|---|---|---|
| Ana | 1, 4, 1 | 0, 4, 0 | +1, 0, +1 | Woman | White |
| Austen | 1, 5, 1 | 0, 6, 0 | +1, -1, +1 | Woman, queer | Asian |
| Bernice | 2, 4, 1 | 1, 3, 2 | +1, +1, -1 | Woman | White, Armenian |
| Clara | 1, 5, 2 | 0, 5, 1 | +1, 0, +1 | Woman | Prefer not to answer |
| Destiny | 4, 2, 3 | 4, 2, 2 | 0, 0, +1 | Man | Black |
| Eclipse | 3, 1, 4 | 3, 3, 3 | 0, -2, +1 | Transmasculine, gender nonconforming | White, Asian/Pacific Islander |
| Eliot | 5, 1, 0 | 4, 2, 0 | +1, -1, 0 | Man | White |
| Emi | 2, 6, 1 | 2, 5, 1 | 0, +1, 0 | Woman | Asian |
| Erin | 1, 1, 4 | 5, 0, 1 | -4, 0, 3 | Transgender, gender nonbinary | White |
| Ethan | 5, 1, 0 | 5, 1, 0 | 0, 0, 0 | Man | White |
| Fisher | 5, 3, 4 | 4, 2, 4 | +1, +1, 0 | Transmasculine | White |
| Godric | 4, 2, 2 | 5, 1, 2 | -1, +1, 0 | Man, cisgender | White, Asian |
| Heather | 2, 4, 0 | 3, 4, 0 | -1, 0, 0 | Woman | White |
| Izzy | 1, 3, 3 | 3, 1, 1 | -2, +2, +2 | Gender nonbinary | White |
| Jaycee | 3, 3, 4 | 2, 5, 3 | +1, -2, +1 | Woman, gender fluid | Hispanic/Latinx |
| Kacey | 3, 3, 6 | 2, 4, 3 | +1, -1, +3 | Woman | White, Asian |
| Kate | 1, 4, 2 | 0, 4, 1 | +1, 0, +1 | Woman, cisgender | White |
| Leo | 6, 0, 0 | 6, 0, 0 | 0, 0, 0 | Man | White |
| Marie | 0, 6, 0 | 1, 5, 0 | -1, +1, 0 | Woman | White |
| Michael | 4, 1, 0 | 5, 0, 0 | -1, +1, 0 | Man | Asian |
| Monica | 2, 3, 6 | 1, 5, 2 | +1, -2, +4 | Gender nonbinary, agender | White |
| Phoebe | 1, 4, 3 | 0, 5, 2 | +1, -1, +1 | Woman | White |
| Rex | 4, 1, 0 | 5, 1, 0 | -1, 0, 0 | Man | Black |
| Sarah | 2, 6, 0 | 1, 6, 0 | +1, 0, 0 | Woman | White |
| Steven | 4, 2, 4 | 4, 1, 5 | 0, +1, -1 | Man | White, Asian |
| Wood | 1, 5, 0 | 1, 5, 0 | 0, 0, 0 | Woman | Asian |

### B. Sources of Gender Appraisal Discrepancies (RQ2)

Table 2 presents student participants' self- and reflected appraisals of masculinity, femininity, and androgyny on the gradational gender metric, along with the calculated appraisal discrepancies. Of the 26 interviewees, 23 reported at least one appraisal discrepancy. Specifically, 19 reported a masculinity discrepancy, 16 reported a femininity discrepancy, and 14 reported an androgyny discrepancy. Below, we summarize two central themes that emerged regarding students' reasonings about the sources of their gender appraisal discrepancies: (1) not feeling personally known by others, and (2) believing that others held different definitions of masculinity, femininity, and androgyny. Notably, some interviewees commented that both factors were particularly pronounced in their physics classes, often more so than in other academic environments. Students attributed this to both the demographic composition and the pedagogical structure of their physics classes.

**1. Not Feeling Known**

When asked why others might perceive their gender expression differently than they did, 16 of the 26 interviewees reasoned that others might not know them well on a personal level. They commented that those who knew them more intimately would be more likely to rate them in line with their self-perceptions. For example, Kate reported appraisal discrepancies for both masculinity and femininity and explained

> Kate: *I think my friends and family probably view me closer to how I view myself—vs. strangers who I don't know very well wouldn't really know that, because it's kind of superficial, just by looking at me and limited interaction*.

Similar to Kate, other interviewees who considered how well others knew them often expressed a belief that, lacking information about their personality, others might draw conclusions about their gender based solely on externally observable information. For example, while explaining how he made decisions about his reflected appraisal ratings, Michael noted that

> Michael: *...the people in my [physics] class, all they have is the visual aspect of me and what I wear. But they don't know who I am, my background, how I act, what I think. And so that's all very random. But once someone gets to know me, the answer definitely will change then.*

Similar to Kate, Michael cited deeper knowledge of his behavior and personality as important factors influencing how others were likely to interpret his gender. He rated himself as more feminine and less masculine than how he believed others would rate him. These discrepancies in masculinity and femininity, along with his explanation of the ratings, suggest there are aspects of his personality he views as feminine but that he believes his peers may not generally see.

Michael's comment also touched on another element that was common to several interviewees' explanations of their reflected appraisals: some students felt that it was common to not know their peers well in physics, often to a greater extent than in other classes. Some students connected these perceptions to the demographics of their physics class. For example, Austen felt that in her physics class, peers may perceive her as more feminine than she perceived herself. She explained that

> Austen: *...because they're all guys I feel like they don't think about it that hard. They're just like, 'okay, she seems to present mostly in that way. She's probably just feminine,' you know.*

Austen described her physics class as dominated by men and felt that they may not "*think about it that hard*" when interpreting her gender expression. She suggested that her classmates, not knowing her well, may make assumptions about her femininity based on the way she outwardly presents.

Other students suggested that the lack of opportunities to get to know one another was also related to the instructional practices in the class. For example, Erin commented that they suspected people in their English class would be more likely to rate them in alignment with how they saw themselves than people in physics and explained that

> Erin: *...a lot of the English classes I've been in have specifically revolved around group discussions about how to take your personal experiences into whatever works you're developing…In physics, we did introductions like 'what's your name? What's your major? And what are you looking forward to in physics this term?' And that was the only group discussion we had before 'all right, let's start doing all these random problems.'*

Erin suggested that structural elements of the physics class contributed to their lack of connection with peers. In their experience, fostering personal connections seemed either not to have been prioritized or not to have come as naturally in physics as in English. Unlike in English, where personal experiences were an integral part of "*whatever works you're developing,*" in physics, the problems were "*random*" and marked the end of personal discussion. Because of this, Erin expressed, they had fewer opportunities to connect with peers in physics. This contributed to Erin's belief that their appraisal discrepancies would be larger in their physics class than their English class.

### 2. Differing Understandings

In addition to not feeling known by others, some students (8/26) also attributed appraisal discrepancies to the feeling that others may have differing definitions or perceptions of masculinity, femininity, and androgyny than themselves. Both students' comments and their use of the gradational gender metric suggested they were hesitant to assume that others had non-stereotypical views of gender. For example, Heather rated herself as less masculine than she believed her peers would rate her, and she explained

> Heather: *I think I'm a very abrasive person. I tend to be very blunt ... and I think that's seen as a more masculine thing. So it's one of my personality traits that I think makes people see me as more masculine—versus me myself, I don't necessarily view that as entirely a masculine thing*.

Heather's masculinity discrepancy appeared to reflect a partial rejection of masculine stereotypes, of which she suspected others in general would be more accepting. She expressed awareness that others might construe her bluntness as masculine but suggested that this did not influence how she viewed her own masculinity.

Moreover, our interviews suggested that the perception that others might hold stereotypical views of gender was especially pronounced in physics learning environments. Seven interviewees expressed that they were more hesitant in physics than in non-STEM or other general settings to assume that their peers would make nuanced gender appraisals. For example, Steven's self-appraisal of femininity was higher than his reflected appraisal, and he explained

> Steven: *[Physics] is a male-dominated class. And at least in my experience, the women in my life have been way more open minded than the men, so I feel like even without knowing me, they would probably put me still not fully masculine, and still not fully feminine anywhere on there.*

Steven explained his appraisal discrepancy by connecting his observation of the male-dominated environment of his physics class with his perception, based on personal experience, that women tend to be more "open-minded" than men. Offering another perspective, Clara, who rated her own masculinity and androgyny higher than she believed her peers in physics would rate her, linked her appraisal discrepancy to the norms and structure in her physics class. When the interviewer asked if her answers would change depending on which class she was considering, Clara drew a comparison between her experiences in physics and liberal arts classes, saying

> Clara: *I think the people in physics class are more likely to perceive me as someone much more feminine than someone in like a Liberal arts class like an education class…because in education we're specifically taught to not make assumptions about people…In that kind of environment…anything goes basically—not like you could say anything, but it's very unbiased when you 1st walk in.*

Clara perceived her education classes to create accepting environments. For her, this observation contributed to a context-dependent appraisal discrepancy. Clara suggested that her peers in education were less likely to overemphasize her femininity because they have been "*specifically taught not to make assumptions about people.*" Clara further explained that her comments did not reflect differing beliefs about the people in physics and education, but rather differences in how classes in the two disciplines, in her experience, were structured. She continued

> Clara: *…we don't talk that much in STEM classes…[Physics] doesn't really allow for that kind of communication…So it will automatically make people do more assumptions than in an education class where you walk in and you're immediately thrown into a conversation about getting to know you, and then [other students] will learn what you perceive yourself as, whatever it is. In STEM it's like, 'hi, okay. 1st impression. Now, therefore, this.' There isn't really much wiggle room.*

.

Clara expressed that she had comparatively fewer opportunities in physics to form interpersonal connections with other students. This influenced her physics-specific reflected appraisals, as she implied that her physics peers had limited opportunity to "*learn what [she] perceived [her]self as*."

### C. Connections between appraisal discrepancies and sense of belonging in physics (RQ3)

In this section, we present our qualitative findings on how self-reflected gender appraisal discrepancies and sense of belonging may be related. We report two primary connections: (1) Most interviewees described feeling pressure to alter their gender expression in the physics learning environment; (2) Both appraisal discrepancies and reduced sense of belonging were linked to students' sense of not being personally known by peers in their physics classes.

#### 1. Tensions associated with pressure to alter gender expression in physics

In our interviews, 14 out of 26 students reported feeling pressure to alter their gender expression in their physics classes. Four women (one also identifying as queer) reported occasionally feeling pressure to accentuate their "feminine" characteristics, while a larger proportion of interviewees (12/26) described feeling a need to present more "masculinely" in physics. Among these 12 students, 8 were women (one of whom also identified as genderqueer and one of whom also identified as genderfluid), 1 was a man, 2 were transmasculine, and 1 identified as agender. For example, Jaycee, who identified as a genderfluid woman, noted that

> Jaycee: *…when I'm in physics, I do like being a little more masculine, but I think that's because I feel like people will listen to me a little better. And so, I know when I'm getting ready, and I know that I have a physics class, or I have a math class or something, I'll try to dress a little more masculine…*

Jaycee sought to accentuate her masculinity in physics contexts because she felt that doing so helped her to garner recognition from peers. Other women spoke similarly about presenting themselves—both in appearance and behavior—in ways they considered masculine. This experience, though, was not exclusive to women. Fisher, a transmasculine student, shared that

> Fisher: *I don't want to act too flamboyant in certain spaces. And usually flamboyance is feminine…In physics class there's lots of times where it's like, 'okay, answer questions with the people around you,' and it's kind of like: 'present masculine. Act like I know what I'm doing. And just keep it very kind of formal.'*

Fisher associated masculinity with the pressure he felt in his physics class to present more formally and convey confidence to his peers. Others similarly noted that they occasionally felt more masculine in their physics classes. Michael, who identified as an Asian man, attributed this to the demographics of his class, explaining that

> Michael: *...a majority of my physics classes have been predominantly guys...I guess I do feel more masculine [there] because there are more guys in the room.*

Godric felt similarly and offered additional rationale. He explained that

> Godric: *...in STEM classes it's been more individual learning. And when I'm by myself, I feel my masculine aspects more.*

For many students, presenting or feeling more "masculine" and less "feminine" was sometimes a semiconscious reaction and sometimes a deliberate response to their perception of the environment in their physics classes. They described these behaviors and sentiments as strategies for fitting into the physics learning environment and gaining recognition or acceptance from their peers.

We note, however, that some students (4/26) occasionally adopted the opposite strategy, presenting themselves in ways they perceived as more stereotypically feminine. Clara, for example, who identified as a white woman, found herself

> Clara: *...being a little hyper-feminine, sometimes in physics...I express myself in a feminine way, but it's also in such like a masculine-dominant area where I feel like I have to overcompensate a lot of the time. Because I'm like, 'I am here.'*

Clara reported that she sometimes accentuated her "feminine" characteristics in physics, which, for her, served as a way of establishing her presence. She wanted to show others that she "was there" amid the predominantly "masculine" presenting students in her class. Unlike the students who expressed a need to emphasize their "masculinity," those who presented themselves in more "feminine" ways did not seek to "fit in." Rather, they sought to stand out. Notably, 3 of the 4 students who reported feeling pressure to present themselves as more "feminine" in physics also reported feeling pressure, at other times, to present themselves as more "masculine."

Among the 14 students who expressed feeling pressure to alter their gender expression in physics classes, 13 explicitly connected these experiences to feelings of discomfort. These experiences were reported across diverse gender identities. Of these 13 students, 8 identified as women (two of whom also identified as either queer or genderfluid), 1 as a man, 2 as transmasculine, 1 as transgender and nonbinary, and 1 as agender. They described a tension between expressing their authentic selves and conforming to gendered expectations they perceived as necessary in the physics environment. Notably, these experiences of inner conflict were most pronounced among students who felt pressure to present themselves more "masculinely." For example, Destiny, who identified as a man, characterized himself as somewhat feminine, self-appraising his femininity as a 2 out of 6. He explained this during the interview by referring to his emotional inclinations, which he described as stereotypically feminine. In the following exchange, Destiny discussed how social contexts influence how he presents himself. The passage begins with his observation that he naturally adopts what he described as more "feminine" presentation when talking to women:

> Destiny: *Sometimes when I talk to a girl…I typically go like this* [here he softens his tone and raises the pitch of his voice] *because I've noticed that a lot of the time it makes them more comfortable to talk to me.*

The interviewer began to ask Destiny if this was something he did purposefully, but he quickly interjected, drawing a connection to his experience in physics:

> Destiny:*—Not always. It's actually a very unconscious thing...I do the same thing in physics class, too, but I'm less likely to do it, though…because other guys are around, and they see me acting like that. And it makes me feel like…yeah.*

Destiny acknowledged that, like the other students discussed in the preceding section, he felt pressure to change his behavior in physics classes. During the interview, he initially trailed off, seeming to lose the words to describe how that made him feel. When prompted further, however, Destiny offered a more nuanced reflection on the emotional complexity of adjusting his behavior in physics:

> Destiny: *...I think it makes you feel uncomfortable and comfortable at the same time… It's like eating a lemon with sugar on it. It feels bad when I'm not actually my real self, but at the same time it gives you some comfort, too, that people are seeing me in a way that I want to be seen.*

Destiny indicated that he did not entirely want to change the way he presented himself to his peers in physics, but his awareness of the male-dominated social setting made him feel a need to tone down aspects of his personality.

Other students made similar comments about feeling pressure to present in more socially congruous ways in their physics classes—ways they often associated with masculinity. As discussed in the previous section, Jaycee described adopting a more stereotypically masculine presentation in physics to be taken more seriously. She later elaborated on the internal conflict this adjustment created:

> Jaycee: *...It does make me feel a little self-conscious about how people perceive me, but it does make me feel internally more comfortable being in the actual class…[there is] a little conflict for me internally...I feel like either way people are looking. It doesn't really feel exactly comfortable in there.*

Destiny's metaphor of *"eating a lemon with sugar on it"* is echoed in Jaycee's account of presenting herself in more stereotypically masculine ways—something she did not for her own comfort, but to better fit into the physics environment. For Jaycee, presenting in this way made her feel *"internally more comfortable"* but also *"self-conscious."* Like Destiny, Jaycee experienced a tension between how she felt implicitly expected to present in physics and how she ideally wanted to express herself. As she explained, presenting more masculinely made her feel that "*people will listen to me a little better."* This left her feeling both uncomfortable and stuck—that, regardless of how she chose to present herself in physics, *"either way people [were] looking."* Notably, Jaycee later shared that this persistent discomfort ultimately contributed to her decision to drop the class.

Kacey, who identified as a woman, echoed and further elaborated on Jaycee's comments, emphasizing that distancing herself from stereotypically feminine presentation was a strategy she used to gain respect and recognition:

> Kacey: *…in the context of any science or math or something related to engineering, I feel like I have to put up a front. I act less feminine. I dress in a way that makes me feel like people take me more seriously. If I wear a cute pink frilly dress, and I do my hair, and I do my makeup, and I feel really good about myself, I find that people don't take me as seriously because they have a certain opinion of people that dress that way.*

Like Jaycee, Kacey reported feeling a need to present in less stereotypically feminine ways so that she would be "*taken seriously*" by her peers. However, the actions she took to achieve this goal required her to sacrifice presenting herself in ways that made her "*feel really good about [her]self.*" While Jaycee and Kacey's comments above focus on appearance, Marie emphasized that, for her, the pressure to perform in more "masculine" ways extended beyond just her appearance. Marie reflected that in her physics classes

> Marie: *I'm worried that I'm going to be perceived more feminine. I feel like I need to compensate for that by possessing those more masculine characteristics to fit in more…[I want] to project a more aggressive and more confident and assertive kind of quality to try and compensate for how I feel.*

Later in the interview, Marie contrasted her experiences in biology and physics. When the interviewer asked how she thought others in biology might perceive her compared with those in physics, she responded

> Marie: *I feel like probably the more biology centric group would be more likely to judge me how I see myself… because I would be less hesitant to act how I feel I am. In physics and in really male dominated environments, I feel like I have to overcompensate with* *some masculine traits to hold my own. In the biology setting, I wouldn't have to do that as much and could really be how I see myself.*

In elaborating on her feeling of being "*hesitant to act how [she feels she is],*" Marie connected her sense of belonging in physics to the pressure she felt to adjust her gender expression in order to better fit in. For her, tension between how she wanted to express herself and how she felt she was expected to present contributed to an ongoing sense of discomfort and inauthenticity in the physics learning environment.

Overall, most interviewed students reported having experienced pressure to alter their gender expression in the physics learning environment, and many of them reported this as a source of internal conflict. These students appeared to experience gender expression in physics as a paradoxical tension: they could be true to themselves and face potential alienation, or they could compromise authenticity for the sake of conformity and recognition. Such tensions were reported across participants with diverse self-described gender identities, including participants who selected one or more of the following: "man," "woman," "nonbinary," "cisgender," "transgender," and "gender fluid".

### 2. Common Factor: Not Feeling Known

Another potential connection between appraisal discrepancies and sense of belonging in physics was related to the overlapping reasoning students used when discussing the two constructs. In answering the second research question, we showed that students often attributed appraisal discrepancies to the feeling that others, especially their peers in physics, did not know them well. Students also cited this as a key factor contributing to lower sense of belonging, suggesting that both experiences may stem from this shared underlying factor. In general, our student participants felt that they belonged in spaces where they could be authentically themselves, and they often cited interpersonal connections as supportive of that feeling. When asked whether having alignment between their self- and reflected appraisals felt important, Erin, a transgender, nonbinary student, shared

> Erin: *Personally, [that feels] really important. I stress a lot. I'm an anxious person overall, and I have that little voice in the back of my head that's saying 'yeah, people don't view you the way you want to view yourself…it's very important to me to be assured not just that my identity is correct, but also that it's okay to share that with other people.*

When the interviewer then asked Erin how that experience manifested in their physics class, Erin explained

> Erin: *It's contributed a little bit to the mindset of just wanting to get through it…A lot of STEM classes are very scientific and analytical. There's not much room for exploring who we are. So I see a lot of my STEM classes as—if I just sit by myself, do the work, take the* *notes, and do well on the quizzes, I can get through it. And then I can move on to classes and other people that I enjoy.*

Erin did not feel supported in their physics class to share their identity with others, and they connected this experience to the stress of feeling that they were not being perceived by others as they would like to be. These sentiments were not exclusive to people who felt a low sense of belonging in physics. For example, Izzy, who identified as gender nonbinary, described the following experience when they started to feel a greater sense of belonging in their physics class:

> Izzy: *I think last term is where I found the group of people that I work with on physics now, and having them know all the things I like to do, and going to activities outside of class, and them seeing who I am…made me feel a lot better in physics classes—partially because I could just freely interact with them, but also it just felt like there was a group that I was really connected with, who shared my interests and could help me with physics if I needed. I think they're what gives me a sense of belonging in classes, just having those other people.*

For Erin, Izzy, and other interviewees, feeling known and understood by peers played a significant role in fostering a sense of belonging in their physics classes. As discussed earlier, students' appraisal discrepancies were also often related to a perceived lack of being known or understood by others. Collectively, these comments suggest that the formation of interpersonal connections (or lack thereof) may represent a common factor underlying both students' self-reflected gender appraisal discrepancies and their sense of belonging.

## VII. Discussion

In this study, we investigated how students in introductory physics courses engaged with a gradational gender expression metric and examined the relationship between their gender expression and sense of belonging in their physics courses. In our prior quantitative work [28,29], we found that differences between how students perceived their own gender expression and how they believed others perceived them—referred to as self-reflected gender appraisal discrepancies—were correlated with students' sense of belonging. In the present qualitative study, we sought to better understand these correlations by examining 1) how students in introductory physics courses understand the terms "masculinity, femininity, and androgyny" and how they place their self- and reflected appraisals on the gradational gender metric (RQ1), 2) the factors contributing to the differences between self- and reflected appraisals of gender expression (RQ2), and 3) the possible connections between students' gender expression, reported appraisal discrepancies, and sense of belonging in their physics classes (RQ3).

Regarding **RQ1,** we found that students' responses to the gradational gender metric were shaped by both common gender stereotypes and more nuanced, context-dependent understandings of the gender expression constructs. Traits students associated with the gender expression constructs, for example "assertiveness" with masculinity and "empathy" with femininity, often aligned with traits that previous studies have identified as elements of gender stereotypes [95–97]. At the same time, most interviewees characterized gender expression as fluid and mutable, and asserted that masculinity and femininity were not mutually exclusive. These findings are consistent with prior studies outside of physics contexts showing that stereotypes and sociocultural content influence individuals' conceptualizations of masculinity and femininity [98–100].

In response to **RQ2**, we found that students most often attributed gender appraisal discrepancies to two primary sources: 1) that others might not know them well, and 2) that others conceptualized masculinity, femininity, and/or androgyny differently than they did. Specifically, interviewees expressed that when others did not know them well, they might make judgments about gender based on superficial, externally available information, such as appearance. Additionally, interviewees expressed hesitancy to assume that others conceptualized gender in non-stereotypical or non-binary ways. Moreover, some interviewees reported that they were more likely to experience gender appraisal discrepancies in physics than in non-STEM classes. Several students suggested that this resulted from perceived barriers to getting to know their peers in physics and other STEM classes. As one participant, Erin, concisely described, in physics class "*there's not much room for exploring who we are.*" Taken together, these findings suggest that limited opportunities for students to become personally known by one another may be an important contextual factor contributing to gender appraisal discrepancies in physics.

The reported barriers to forming interpersonal connections could potentially be related to broader cultural features of the discipline, as suggested by prior studies of physics environments. In her ethnographic account of high energy physicists, Traweek portrayed physicists' regard for objectivity as a claim to inhabit a "culture of no culture" [101]. More recent research has shown that, in some cases, appeals to objectivity may unintentionally contribute to the maintenance of inequitable systems by discouraging engagement with issues of equity [102–104]. Since Traweek's report, numerous scholars have also explored the so-called "myth of objectivity" and other facets of the physics "non-culture," calling attention to the ways in which they inhibit discussion of socially rooted issues [54,102,105]. In our study, students often contrasted technical, objective discussions in physics to more social or subjective discussions in other disciplines. Some interviewees perceived the lack of opportunities to discuss social issues as a barrier to forming interpersonal relationships with their peers. While additional research is needed, our findings suggest that when social and interpersonal aspects of classroom interaction are not emphasized, students may have fewer opportunities to know one another personally, which may in turn contribute to gender appraisal discrepancies.

For **RQ3**, we identified two potential explanations for the previously observed correlations between positive masculinity/negative femininity appraisal discrepancies and sense of belonging in physics [28,29]. First, many interviewees described altering their behavior or gender expression in physics learning environments, often noting that doing so felt uncomfortable, inauthentic, or misaligned with their sense of self. Second, students often described both appraisal discrepancies and sense of belonging in connection with perceptions that their peers in physics did not know them well.

Among the students who felt pressure to alter their gender expression in physics contexts, experiences of internal conflict and tension were most evident for those who reported pressure to downplay their "feminine" attributes in favor of more "masculine" ones. This may help contextualize the asymmetric pattern observed in our previous quantitative results, which showed that students who perceived their "masculinity" as underrecognized and/or their "femininity" as overemphasized tended to report lower sense of belonging than those who reported no discrepancy or a discrepancy in the opposite direction [28,29]. For many interviewees, the construct of "masculinity" encompassed elements of behavior that they perceived to be effective for garnering recognition and respect in their physics classes. In contrast, students often associated "feminine" expressions with marginalizing experiences such as being ignored or overlooked. From this perspective, the association between lower sense of belonging and positive discrepancies in masculinity (i.e., perceiving one's masculinity as underrecognized), as well as negative discrepancies in femininity (i.e., perceiving one's femininity as overemphasized) may reflect students' perceptions that styles of self-expression associated with femininity are less readily recognized as valuable in physics contexts. This interpretation aligns with prior literature documenting how women in physics contexts distance themselves from stereotypical femininity in order to navigate the gendered disciplinary culture of physics [13–15,72,73]. Our findings further suggest that such dynamics may explain why specific appraisal discrepancies were linked to lower belonging in the prior quantitative study. More broadly, they suggest that appraisal discrepancies may reflect students' experiences of recognition, misrecognition, and pressure surrounding self-expression in physics.

At the same time, interviewees described multiple—and sometimes contrasting—strategies for managing these gendered expectations. While some students reported downplaying femininity and amplifying masculinity in order to gain recognition, others described deliberately asserting their femininity to stand out or resist being overlooked. These approaches resonate with the previously described strategies of "fragmentation" and "multiplicity" through which individuals navigate competing identity expectations by selectively minimizing, emphasizing, or holding together different aspects of themselves across contexts [54,75]. Our findings further suggest that students' sense of belonging in physics may be challenged when they perceive pressure to express their gender in ways that feel inauthentic.

Importantly, we found that pressures to alter one's gender expression and the associated feelings of inauthenticity were reported by students across multiple gender identity groups, building on prior work documenting women's negotiations of femininity in physics [13–15,54] and gender minority students' reports of pressure to conform or conceal aspects of their gender expression in STEM/physics contexts [8,9,16,17,59]. These findings suggest that tensions related to self-expression in physics may be shaped by how individuals' gender expressions align with norms perceived as valued within the discipline. In contexts where competence and recognition are often associated with stereotypically masculine traits, students whose self-expression diverges from these norms may be more likely to experience pressure to conform to them.

As a second potential explanation for **RQ3**, we found that both gender appraisal discrepancies and lower sense of belonging were associated with students' perceptions that their peers in physics class did not know them well. Some students linked the experience of not feeling known to their perception that it was difficult to personally connect with their classmates. This finding aligns with a prior study that reported limited opportunities for students to form sustained peer connections within physics classes [106]. That study, which used a mixed-methods approach involving social network analysis and semi-structured interviewing, found that the number of study group partnerships in the social network of an introductory physics class "decreased substantially as the semester progressed" [106]. Collectively, these results suggest that opportunities for peer connection within physics classes may be limited. Our findings further indicate that such limitations may matter not only for students' general social experiences, but also for how students believe they are perceived by others. In this way, the perceived lack of opportunities for interpersonal connection may help explain both reduced sense of belonging and the emergence of gender appraisal discrepancies.

## VIII. Implications for Instruction and Future Research

The findings discussed above point to several implications for gender-inclusive physics instruction and suggests additional directions for future research. Amid calls to move beyond binary gender metrics [10,77], our results highlight the potential value of gender expression as a complementary lens and gradational gender expression metrics as promising tools for studying student learning experiences in physics. Although students may hold varied understandings of femininity, masculinity, and androgyny, our results suggest that discrepancies between self- and reflected appraisals of gender can encode meaningful perceptions and experiences. Specifically, our results suggest that appraisal discrepancies are context-sensitive and might therefore be used to investigate how students' experiences are shaped by features of their learning environment, such as classroom culture or peer interactions. Additionally, by documenting similar pressures across multiple gender identity groups, this work suggests that pressures related to gender expression in physics are not confined to a single gender identity group. Future studies could more systematically examine which forms of gender expression are perceived as competent, legitimate, or recognizable in physics classrooms, and how these perceptions shape students' participation and sense of belonging.

Our findings concerning **RQ2** and **RQ3** suggest that fostering opportunities for meaningful peer connection may be an important consideration for creating more inclusive physics learning environments. In particular, students' perceptions of being known and valued by peers appeared to be closely tied to both sense of belonging and gender appraisal experiences. This suggests that instructors might seek to design classroom interactions that support not only academic collaboration, but also interpersonal connection. Establishing positive regard among peers may be particularly influential. Prior research has suggested that social caring among peers constructively influences learning by encouraging productive framing and engagement [107,108], and that adolescents who perceive that they are valued and respected by their peers are likely to have stronger motivational beliefs [109]. Instructors can design lessons, build curricula, and influence classroom climate to facilitate the formation of positive relationships among students. Prior research suggests that pedagogical strategies that may support this type of engagement include creating collaborative, context-rich learning opportunities, modelling effective communication strategies, and attending to the physical layout of learning spaces [110–114].

Our interviews also suggest that barriers to peer interaction and connection in physics learning environments may be linked to students' perceptions of safety and recognition. Instructors can seek to address this by making intentional efforts to build inclusive learning environments in which students feel safe to express themselves. For example, instructors may consider implementing ecological-belonging interventions, which have been linked to reduced performance gaps between marginalized students and their dominant group peers [115,116]. Wang and Hazari show that teacher recognition can take a number of explicit and implicit forms, including establishing high expectations, devoting resources, providing affirmations and encouragement, and centering student learning by providing opportunities for active engagement and multiple modes of participation [117]. In our interviews, we observed that some students expressed hesitancy to interact with peers when class activities lacked meaningful context or placed a high emphasis on the correctness of the final answer. Instructors might therefore also consider how they structure problems and activities to support the development of inclusive group interactions and psychological safety during collaborative learning activities.

Universities, departments, and instructors alike might also place more intentional emphasis on developing students' interpersonal and social-emotional competencies, such as empathy, communication, and critical social consciousness. Such competencies are essential for forming interpersonal connections, especially across cultural boundaries, yet several studies have suggested that physics and other STEM majors do not have adequate opportunities to develop them [118–120]. At an institutional or departmental level, efforts to improve such outcomes might include creating more holistic degree requirements to ensure graduates enter the workforce with both the technical skills and the interpersonal and social-emotional competencies needed to have a meaningful, positive impact in their chosen careers. Some universities have approached this goal by developing degree requirements that ask students to engage with issues of difference, power, and oppression within their discipline [121]. At the course level, instructors can consider designing classes or lessons around real-world issues involving physics. For example, some instructors have explored teaching physics concepts through units on sustainability and nuclear weaponry [122,123]. Research on these pedagogies found that they helped structure in-depth discussions of ethics and social responsibility [122,123]. Through such practices, educators may help reduce barriers and social risks that contribute to disconnection and discomfort among physics students.

## IX. Limitations

Although we made efforts to recruit interviewees from a variety of cultural and ethnic backgrounds, the generalizability of our results remains limited by the context in which the research took place. This study was conducted at a large public R1 institution with a predominantly White student population, which may have limited the diversity of perspectives present within the population of physics students. Students' perceptions may also have been shaped by the format of their classes. The introductory physics courses from which we recruited student participants were large, with total enrollment in each course typically around 200. This study is therefore unable to entirely distinguish between class size and elements of the physics culture as sources of the student perceptions discussed above. Future research might seek methods to control for the potentially confounding effects of class size.

## X. Acknowledgements

We thank all students who participated in the study and Dr. Liz Gire and Dr. Patti Hamerski for their constructive feedback on the manuscript.